\documentclass[twocolumn,prd,aps,groupedaddress,showpacs,nofootinbib,preprintnumbers]{revtex4}

\usepackage{amsmath}
\usepackage{amsfonts}
\usepackage{graphicx}
\usepackage{bm}

\def\be{\begin{equation}}
\def\ee{\end{equation}}
\def\ba{\begin{eqnarray}}
\def\ea{\end{eqnarray}}
\def\bs{\begin{subequations}}
\def\es{\end{subequations}}
\def\vr{\varrho}
\def\rmm{\text{m}}
\def\rmr{\text{r}}
\newcommand{\Eq}[1]{(\ref{#1})}

\begin{document}

\title{Tachyon dark energy models: Dynamics and constraints}
\author{Gianluca Calcagni}
\author{Andrew R. Liddle} 
\affiliation{Astronomy Centre, University of Sussex, Brighton BN1 9QH,
United Kingdom}
\date{May 31, 2006}

\begin{abstract}
We explore the dynamics of dark energy models based on a
Dirac--Born--Infeld (DBI) tachyonic action, studying a range of
potentials. We numerically investigate the existence of tracking
behaviour and determine the present-day value of the equation of state
parameter and its running, which are compared with observational
bounds. We find that tachyon models have quite similar phenomenology
to canonical quintessence models. While some potentials can be
selected amongst many possibilities and fine-tuned to give viable
scenarios, there is no apparent advantage in choosing a DBI scalar
field instead of a Klein--Gordon one.
\end{abstract}

\pacs{98.80.Cq}
\preprint{astro-ph/0606003}

\maketitle


\section{Introduction}

Most dark energy modelling using scalar fields has followed the
quintessence paradigm of a slowly rolling canonical scalar field.
However, there has been increasing interest in loosening the
assumption of a canonical kinetic term. In its most general form, this
idea is known as k-essence \cite{AMS}. A more specific choice is the
`tachyon' \cite{gib02}, which can be viewed as a special case of
k-essence models with Dirac--Born--Infeld (DBI) action
\cite{chi03}. This kind of scalar field is motivated by string theory
as the negative-mass mode of the open string perturbative spectrum,
though its use in the dark energy sector is primarily
phenomenological. One goal of such studies is to
investigate whether there are any distinctive signatures of
non-canonical actions available to be probed by observations.
For a recent comprehensive review of dark energy
dynamics, see Ref.~\cite{CST}.

Tachyon dark energy has been explored by many authors, for example
Refs.~\cite{CGJP,HL,BJP,GST,CGST,car06}.\footnote{Here we do not
consider the tachyon either as a dark matter candidate \cite{SW,PC} or
as the inflaton field.}  Two papers are particularly closely related
to the present work. Bagla \emph{et al.}~\cite{BJP} focussed on two
specific choices of tachyon potential, and carried out numerical
analysis of the cosmological evolution in order to constrain them
against supernova data and the growth rate of large-scale
structure. Copeland \emph{et al.}~\cite{CGST} studied a wider range of
potentials, concentrating mainly on analytical inspection of attractor
behaviour and the critical point structure without making comparison
to specific observations. In this paper, we aim to merge some of the
positive features of each analysis, by studying a wide range of
potentials and testing them directly against current observational
constraints as given in Ref.~\cite{WM}.

The mechanism of slow rolling is the key ingredient in order to get an
accelerating evolution driven by a scalar field. Since the DBI action
can be expanded to match the Klein--Gordon one in this regime, one
does not expect to find radically different features from the
traditional quintessence models. However, we point out that, as
compared to canonical quintessence, tachyon models require more
fine-tuning to agree with observations. This is consistent with the
properties of the slow-roll correspondence between the tachyon and an
ordinary scalar \cite{ben02,GKMP}.


\section{Tachyon dark energy models}


\subsection{Equations of motion}

We assume a four-dimensional, spatially-flat
Friedmann--Robertson--Walker Universe filled by dust matter
(subscript `m'), radiation (`r') and a minimally coupled homogeneous
DBI tachyon $T$ with potential $V(T)$ and dimension $E^{-1}$. For a
perfect fluid $n$ with energy density $\rho_n$ and pressure $p_n$, the
barotropic index is $w_n\equiv p_n/\rho_n$. Each fluid component
satisfies a continuity equation
\be
\dot\rho_n+3H\rho_n(1+w_n)=0 \,,
\ee
with $w_{\rm m}=0$ and $w_{\rm r}=1/3$. Here, a dot is derivation with
respect to synchronous time and $H\equiv \dot a/a$ is the Hubble
parameter defined in terms of the scale factor $a(t)$. In the
following a subscript 0 will denote quantities evaluated today (at
$t_0$), when $a(t_0)=a_0=1$. Defining the critical density today as
$\rho_{{\rm c}0}\equiv 3H_0^2/(8\pi G)$ and assuming that gravity obeys
an Einstein--Hilbert action, the Friedmann equation reads 
\ba
H^2   &=& \vr_{\rmm}+\vr_{\rmr}+\vr_T\,,\\
\vr_{\rmm} &=& \Omega_{\rmm,0}a^{-3}\,,\\
\vr_{\rmr} &=& \Omega_{\rmr,0}a^{-4}\,,\\
\vr_T &=& \frac{U(T)}{\sqrt{1-\dot T^2}}\,,
\ea
where $\vr_n\equiv\rho_n/\rho_{c0}$, $U\equiv V/\rho_{c0}$, and the
time coordinate has been rescaled as $t\to t/H_0$ so that $H=1$
today. The tachyon is dimensionless in these units. Note that the
$\vr_n$ are the densities normalized to the \emph{present} value of
the critical density, and are not the density parameters. The present
density parameters have values $\Omega_{\rmm,0}\approx 0.24$ \cite{spe06}
and $\Omega_{\rmr,0}\approx 8\times 10^{-5}$.\footnote{As input in our
numerical code we have chosen $\Omega_{\rmm,0}= 0.25$. All results are
unaffected by small changes in the matter density.}

The tachyon equation of motion is \cite{FT}
\be
\frac{\ddot T}{1-\dot T^2}+3H\dot T+ \frac{U_{,T}}{U}=0,
\ee
where $U_{,T}\equiv d U/d T$. Since $p_T=-V\sqrt{1-\dot T^2}$,
the barotropic index for the tachyon is
\be
w_T=\dot T^2-1,
\ee
which can vary only between $-1$ and $0$ in order for the action to be
well defined. When the scalar field slowly rolls down its potential,
$|\dot T|\ll 1$, it behaves like an effective cosmological constant,
$w_T \approx -1$. From now on we drop the subscript `$T$' on $w_T$.

Since the cosmological evolution spans many orders of magnitude in
synchronous time, for numerical work it is convenient to switch to the
number of $e$-foldings $N\equiv \ln a/a_0=\ln a$ as the evolution
parameter, so that 
\be
\vr_{\rmm} = \Omega_{\rmm,0}e^{-3N} \,, \qquad \vr_{\rmr} =
\Omega_{\rmr,0}e^{-4N}\,. 
\ee
Derivatives with respect to $N$ will be denoted by primes, so that
$\dot Q=HQ'$ and $\ddot Q=H^2(Q''-\epsilon Q')$ for any quantity $Q$,
where $\epsilon\equiv -\dot H/H^2=-(H^2)'/(2H^2)$. With the compact
notation $x\equiv H^2$, $\vr\equiv \vr_\rmm+\vr_\rmr$, the Friedmann
equation becomes 
\be\label{fr}
x=\vr+\frac{U}{\sqrt{1-x T'^2}}\,.
\ee
The equation of motion for the tachyon is
\be
\frac{x(T''-\epsilon
T')}{1-xT'^2}+3xT'+\frac{U_{,T}}{U}=0\,,\label{eomt}
\ee
where
\be
\epsilon = -\frac{x'}{2x}\,,\qquad
x' = -3\vr_\rmm-4\vr_\rmr-\frac{3xUT'^2}{\sqrt{1-xT'^2}}\,.
\ee
{}From Eq.~\Eq{eomt} one finds
\be
T''=-\frac1x
\left[\frac{x'T'}{2}+(1-xT'^2)
\left(3xT'+\frac{U_{,T}}{U}\right)\right]\,.   
\ee
Mapping between the number of $e$-foldings and the redshift
$z\equiv a_0/a-1$, we note that at big bang nucleosynthesis (BBN) $N_{\rm
BBN}\approx-20$ ($z\approx 10^9$), at matter--radiation equality
$N_{\rm eq}\approx -8$ ($z\approx 3200$), and at recombination $N_{\rm
rec}\approx -7$ ($z\approx 1100$).

As regards the initial conditions at early times for the dynamical
equations, we can consider two qualitative cases. In the first one,
the scalar field starts rolling down very slowly, $xT'^2\ll 1$. The
Friedmann equation becomes linear in $x$ and, during radiation
domination, $\vr\gg U$ and $x\approx \vr$. In such a picture of the
events, the initial condition is a radiation-dominated Universe in
which a dynamical cosmological constant is negligible. Later on,
$\vr_T$ increases relative to $\vr_{\rm r}$, eventually dominating at
low redshift, while $w$ stays negative and varies, perhaps
non-monotonically, from around $-1$ to $w_0<0$. The parameters of the
model and the shape of the potential can be adjusted so that the
actual value of $w_0$ is compatible with supernov\ae\ data.

In the second case $xT'^2\lesssim 1$, that is, $T'$ is large enough
to compensate, but not override, $x\gg 1$. The tachyon starts from a
dust regime and ends up again with a suitable negative $w_0$. This is
precisely the situation studied numerically in Ref.~\cite{CGST} for
$U\propto T^{-1}$.

Below, we shall focus mainly on the first case, which encodes all the
relevant features of the models.


\subsection{Tracking and creeping regimes}

Quintessence behaviour typically falls into one of two classes, named
as \emph{tracking} and \emph{creeping} in Refs.~\cite{SWZ,WCOS} (see also Refs.~\cite{CL,lin06} which refer to the more general late-time
behaviour in these scenarios as \emph{freezing} and \emph{thawing}). The same kind of evolution also occurs in the tachyon case.

Tracking behaviour occurs when the potential has an attractor as a
response to a cosmological fluid, which renders the final outcome (at
fixed fluid density) independent of the initial conditions. Many
potentials support tracking behaviour, which takes place provided the
initial dark energy density is not too low (otherwise the scalar field
cannot dominate by the present).  A tracking field slows down near the
present as it starts to feel the friction induced by its own dominance
of the energy density, and hence has equation of state reducing with
time ($w'<0$). Note that an asymptotic solution, in which radiation
and matter are negligible and dark energy is the only component, may
be an attractor but is not a tracker according to the above
definition; tracking behaviour is induced by another fluid component.

A creeping field is one which sits static at low energy density until
the density of other materials drops low enough for it to become
dynamical and start to move at the present epoch. Creeping dark energy
typically does not make predictions independent of initial conditions,
but can be characterized by the equation of state increasing away from
$-1$ at the present epoch as the field starts to move ($w'>0$).

Potentials with tracking regimes also feature creeping behaviour for
low enough initial density, which can lead to different predictions
from the same potential. Creepers can be seen mainly as models which have passed the
attractor and are frozen until they reach it, although creeping behaviour can also be found for many
potentials which do not support \emph{observable} tracking.\footnote{Whenever there is an attractor (either dust or de Sitter) in the phase-space plane, the asymptotic solution of course does not depend on the initial conditions. However, if these are `too far away' from the attractor, it may not be possible to get close enough to a tracker/attractor before matter has reached its present density.}

In order to regard the tachyon as a model of dynamical dark energy, it
is useful to parametrize the barotropic index as a function of the
scale factor:
\be
\label{wa}
w=w_0+w_a(a-1),
\ee
so that $w=w_0$ today. Its value can be found by solving the Friedmann
equation today for $T'_0=T'(0)$, getting  from Eq.~\Eq{fr}
\be
w_0=(T'_0)^2-1=-\left(\frac{U_0}{1-\varrho_0}\right)^2.
\ee
{}From Eq.~\Eq{eomt}, one has that
\ba
w_a &=&e^{-N} w',\\
w' &=& 2wT'\left(3xT'+\frac{U_{,T}}U\right)\label{wpr1}\,.
\ea
Then one has $w'<0$ when either
$T'<0$ and $3xT'+U_{,T}/U<0$ (for inverse power-law potentials, $U\sim
T^{-\beta}$, the latter condition is $3xTT'-\beta<0$, true when
$T>0$), or $T'>0$ and $3xT'+U_{,T}/U>0$.

The transition between a tracker and a creeper can be defined by
imposing an initial condition for $T$ and $T'$ so that $w'_0=0$.
Genuinely non-tracking models are problematic as they do not make
definite predictions, in the sense that there is strong dependence on
initial conditions. This makes them rather unattractive, although by
no means ruling them out. Indeed, it makes some of them more viable
than trackers in relation to experimental bounds, as we shall see.


\subsection{Choice for the tachyon potential}\label{list}

As for the tachyon self-interaction, there are a number of models
which one can consider, some being motivated by nonperturbative string
theory and others purely by phenomenology. We review the
classification of Ref.~\cite{CGST}, to which we refer for further
details and references. For each case, past results are summarized. In
the following, $U_c$ is a normalization constant.

\begin{enumerate}
\item $U=U_c T^{-\beta}$, $\beta<0$, $\epsilon\propto T^{\beta-2}\to
\infty$ as $T\to 0$. This model is affected by instabilities \cite{FKS}.

\item $U=U_cT^{-\beta}$, $0<\beta<2$, $\epsilon\propto T^{\beta-2}\to
0$ as $T\to +\infty$ [asymptotically de Sitter (dS)]. In order to get
viable cosmologies, $U_c$ does not need to be fine-tuned since it is
not affected by the super-Planckian problem of the inverse quadratic
potential. In general there is a stable late-time attractor
\cite{AF}. In Ref.~\cite{CGST} the case $\beta=1$ was considered as a
numerical example.

\item $U=U_c T^{-2}$, $\epsilon={\rm constant}$. This is the potential
associated to the exact power-law solution $a=t^p$ \cite{BJP,AF,pad02}. One
has to fine-tune $U_c$ in order to get sufficient acceleration
today. A phase-space analysis in Refs.~\cite{AL,CGST} confirms that it
is difficult for this model to explain dark energy, since the only
late-time attractor in the presence of matter or radiation has
$\varrho_T/x\to 0$.

\item $U=U_c T^{-\beta}$, $\beta>2$, $\epsilon\propto T^{\beta-2}\to
\infty$ as $T\to +\infty$. This potential has not been studied
numerically previously, but Ref.~\cite{AF} showed that it has a dust
attractor. The authors of Ref.~\cite{CGST} also argued that it behaves
as in model~7 below.

\item $U=U_c\exp(1/\mu T)$, $\mu>0$, $\epsilon\propto \exp(-1/\mu
T)/T^2\to 0$ as $T\to +\infty$. This gives an asymptotic dS solution
with effective cosmological constant given by $U_c$. This potential
has been considered also for Klein--Gordon quintessence
\cite{ZWS,SWZ}, and it should have properties similar to model 2
\cite{CGST}.

\item $U=U_c\exp(\mu^2T^2)$, $\mu>0$, $\epsilon \propto
T^2\exp(-\mu^2T^2)\to 0$ as $T\to 0$. This potential arises in KKLT
setups \cite{KKLT} for massive scalar modes on the $D$-brane \cite{GST}. The field
oscillates around $T=0$ with effective cosmological constant given by
$U_c$, and can give viable scenarios~\cite{CGST}.

\item $U=U_c\exp(-\mu T)$, $\mu>0$, $\epsilon\propto \exp(\mu
T)\to\infty$ as $T\to +\infty$. This potential arises in $D$--$\bar D$
systems of coincident branes with a real tachyonic mode, and was
studied in Ref.~\cite{BJP}. The authors of Ref.~\cite{CGST} found it
to have a stable dust attractor after a period of acceleration. They
suggested the possibility that we are living during this transient
regime. Note that this potential is a large-field approximation of
$U=U_c/\cosh(\mu T)$ \cite{LLM}.

\item $U=U_c\exp(-\mu^2T^2)$, $\mu>0$, $\epsilon \propto
T^2\exp(\mu^2T^2)\to \infty$ as $T\to +\infty$. In Ref.~\cite{CGST} it 
was argued that its predictions are similar to model 7.
\end{enumerate}
These potentials offer a rather comprehensive range of generic
behaviours: one reaches a cosmological constant regime at infinite
time [$U\propto \exp(T^{-1})$], one at finite time ($U\propto \exp
T^2$), while in other situations the potential asymptotically
vanishes.

The structure of analytic asymptotic solutions of the equations of
motion for inverse power-law and exponential potentials is presented
in the Appendix.


\section{Numerical analysis}

The equations of motion are integrated forward in time from before
matter--radiation equality ($N_{\rm i}\approx -10$).\footnote{Although
the problem is defined by boundary conditions at the present time, we
do not attempt to integrate backwards from them (for instance by
choosing $T_0$ freely while $T'_0$ is constrained by the Friedmann
equation at $N=0$). This is because in many cases we are dealing with
situations having strong attractors, which become repellers in
backwards integration leading to rapid growth of numerical
instabilities. Instead, we integrate forwards using a shooting method
to adjust the initial conditions to obtain the right present
properties.}  Having fixed the initial values $T_{\rm i}$ and $T'_{\rm
i}$, our numerical code adjusts the normalization of the potential to
yield the solutions which satisfy the boundary conditions today
($x_0=1$, matter and radiation energy densities equal to $\Omega_{\rm
m,0}$ and $\Omega_{\rm r,0}$, respectively). In cases where a tracker
behaviour is present, one would expect the present values of these to
be independent from the initial conditions to a good approximation.

We choose $T'_{\rm i}=0$ as initial condition; an arbitrary
non-vanishing value could typically only have arisen if at very early
times the tachyon had an unacceptably high velocity.  The initial
phase of approach to a tracker is artificial, as the real cosmological
initial conditions were presumably laid down at an earlier epoch than
those of our code. Anyway, it will be sufficient to capture the main
features of these models, although one might devise some physical
situations which do predict $U\ll \varrho$ and $0<T'\ll 1$ as initial
conditions.

In the following we inspect case by case the models listed in
Sec.~\ref{list}. Since case 1 is unstable, we start with inverse
power-law potentials.

\begin{figure}[t]
\includegraphics[width=8cm]{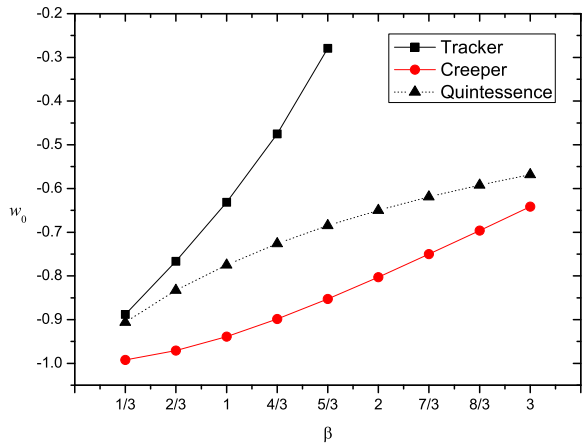}
\caption{\label{fig1} Present value of the barotropic index $w$ for
tachyon tracking solutions (squares) and for creeping solutions for
$T_{\rm i}=1$ (circles) with inverse power-law potentials $U\propto
T^{-\beta}$. For comparison the result for tracking canonical
quintessence is also shown (triangles).}
\end{figure}


\subsection{$U=U_cT^{-\beta}$}\label{power}

These potentials lead asymptotically to dS for $\beta<2$ and to a dust
attractor for $\beta \geq 2$. The first case is the most viable for
obvious reasons, and we have verified that for $T_{\rm i}<0.1$ and
$T_{\rm i}'=0$ the numerical solutions with $\beta<2$ lie on the
asymptotically dS attractor, while for $\beta\geq 2$ and the same
initial conditions it is not possible to achieve a cosmology
compatible with observations (i.e., $x_0=1$). For $T_{\rm i}\gtrsim
0.1$, one goes further and further away from the attractor and enters
a creeping regime, in the sense that the evolution depends on the
choice for $T_{\rm i}$. In Fig.~\ref{fig1} the value of $w_0$ is shown
for the tracker solution and for a particular creeping solution with
$T_{\rm i}=1$. In general, creepers mimic a cosmological constant more
closely than trackers, and, in particular, only a creeping solution is
available for $\beta \geq 2$. For comparison we also show the equation
of state from tracking canonical quintessence
models,\footnote{Canonical quintessence follows a Klein--Gordon
equation $\ddot\phi+3H\dot\phi+V_{,\phi}=0$. The tracking regime is
reached from small initial values of the scalar field $\phi$
($\phi_{\rm i}\ll 1$ in Planck mass units).} which for the same
$\beta$ is always more negative than that of the tracking tachyons.

These results can be understood by recalling that in the slow-roll
approximation there is a mapping between the DBI scalar and the
Klein--Gordon one \cite{ben02,GKMP}. The tachyon energy density is
$\vr_T = U(T)/\sqrt{1-\dot T^2}\approx U(T)+U(T)\dot T^2/2$, and after
a field redefinition $\phi\equiv \int\!\sqrt{U(T)}\,dT$ one obtains a
canonical action for a scalar $\phi$ with potential $W(\phi)\equiv
U[T(\phi)]$. For $U(T)\sim T^{-\beta}$, one has
\be
W(\phi)\sim \phi^{-\tilde\beta} \quad , \quad
\tilde\beta=\frac{2\beta}{2-\beta} \,.  
\ee
Hence the tachyon tracker for $0<\beta<2$ is dual to the tracking
quintessence solution $\tilde\beta>0$. From Fig.~\ref{fig1}, one can
see that any dual pair $(\beta,\tilde\beta)$ corresponds to almost the
same index $w_0$ --- see in particular the pairs $(2/3,1)$ and
$(1,2)$.

\begin{figure}[t]
\includegraphics[width=8cm]{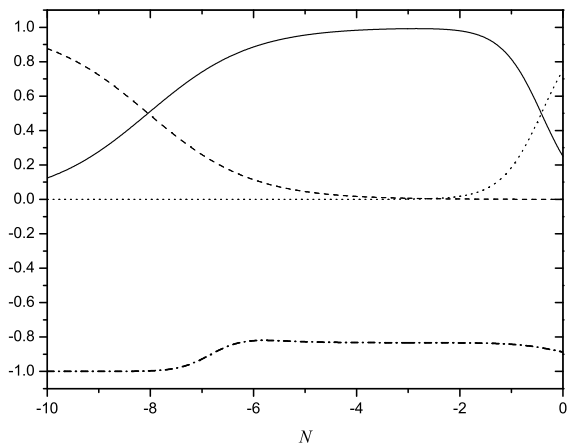}
\caption{\label{fig2} Evolution of the density parameters and equation
of state as a function of $N$ for the inverse power-law tachyon model
with $\beta=1/3$. Solid line: $\varrho_{\rm m}/x$; dashed line:
$\varrho_{\rm r}/x$; dotted line: $\varrho_T/x$. The dot-dashed line
is the barotropic index $w$.}
\end{figure}

The attractor evolution of the fluid components is shown in
Fig.~\ref{fig2} together with $w(N)$, for the particular case
$\beta=1/3$. The behaviours of these quantities for other values of
$\beta$ are all similar. Note that the barotropic index is
$w\approx-1$ until matter domination, increases up to $w\approx -0.8$
at redshift $z\sim 10^2$, and then decreases towards $-1$. This is
consistent with the `freezing' behaviour and its limit of
applicability as classified by Ref.~\cite{CL}.

For the creeping solutions (not shown here) the index $w$ is very
close to $-1$ up to very late times, when it starts deviating to a
softer equation of state.

In this and all other examples, the onset of dark energy lies in the
interval $-1<N<0$, that is, for redshift $z<1.7$. Such late domination
by dark energy is essential to prevent an excessive suppression of
structure formation growth, see e.g.~Ref.~\cite{BJP}.

\begin{figure}[t]
\includegraphics[width=8cm]{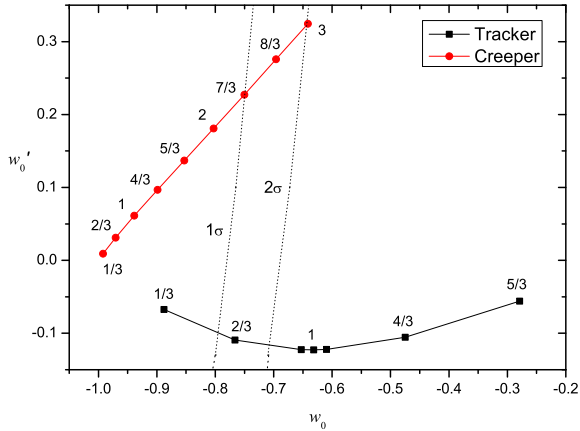}
\caption{\label{fig3} Numerical points in the $w_0$-$w'_0$ plane for
tracking solutions (squares) and creeping solutions for $T_{\rm i}=1$
(circles) with inverse power-law potentials $U\propto T^{-\beta}$. The
value of $\beta$ is shown at each point, while the dotted lines are
the $1\sigma$ and $2\sigma$ likelihood bounds \cite{WM}. The
unlabelled points on the left and right of $\beta=1$ in the tracking
curve are $\beta=0.95$ and $\beta=1.05$, respectively.}
\end{figure}

In Fig.~\ref{fig3} the points predicted by the same models in the
$w_0$--$w'_0$ plane are shown together with the $1\sigma$ and
$2\sigma$ likelihood contour bounds of Ref.~\cite{WM}. These are based on the first-year SNLS data set
\cite{ast05} and SDSS \cite{eis05}, WMAP3 \cite{spe06}, and 2dF
\cite{ver02,haw02} experiments.  Note that there is a minimum value
for $w'_0$ around $\beta\approx 0.99$; we have checked that its
closeness to 1 is an accident of the particular $\Omega_{\rm m,0}$ used.

The conclusion is that there is tracking behaviour for $\beta<2$, and
if $T_{\rm i}$ is not too large this operates and gives a well-defined
prediction for $w_0$, which is however in disagreement with
observations unless $\beta$ is quite small. The predicted $w_0$ moves
closer to $-1$ as the potential flattens, but only once it gets down
to about $\beta \lesssim 0.8$ do we start to see compatibility with
the $2\sigma$-level bounds. This is even flatter than the equivalent
constraint on quintessence, $\beta \lesssim 4/3$. Hence these models
are not very pleasant from either a theoretical or experimental point
of view. One has to impose $\beta \lesssim 1$ or fine-tune the initial
conditions. Relaxing either assumption results in incompatibility with
observations. The creeping solution with $T_{\rm i}=1$ (upper curve)
gives an example of observationally viable model for $\beta \leq 3$,
but at the cost of fine-tuned initial conditions.

One can also make $T'_{\rm i}$ big enough that the initial $w$ is
close to zero rather than $-1$. Then it dips to $-1$ and increases to
join the tracker, eventually giving exactly the same evolution as if
one had chosen $T'_{\rm i} = 0$ at the start. This is exactly the same
sequence of epochs that are seen in quintessence models \cite{BMR}; if
the initial velocity is set high, then the trajectory `overshoots' the
tracker, comes to rest at lower potential energy for a while, and
eventually rejoins the tracker from below.

We can compare the numerical output for $T_0$ with the asymptotic
solutions derived in the Appendix, specifically Eq.~\Eq{to}. With
$C=0.045$, the ratio $|T_0^{\rm num}-T_0^{\rm theor}|/T_0^{\rm num}$
is always less than 2\%. As explained in the Appendix, the asymptotic
analytic solution relies on a polynomial rather than monomial
potential, and in general they will predict a different variation of
$w$. Moreover, the values of $p$ from Eq.~\Eq{p} are rather small,
corresponding to $-2.5\lesssim N_*\lesssim -0.5$, where $N_*$ is the
big bang event; i.e.~the asymptotic solution which matches the present
evolution has a big bang in the very recent past, and hence can only
have become a reasonable approximation very recently. The solution
properly approaches the attractor only for large positive $N$, when
the contributions of radiation and dust matter are
negligible. Nevertheless it appears to give a good estimate of $w_0$,
though not $w_0'$.


\subsection{$U=U_c\exp(1/\mu T)$}\label{exp-1}

This model has a non-zero vacuum energy $U_c$ as \mbox{$T \rightarrow
\infty$}, which will be achieved asymptotically. It exhibits some
features similar to the inverse power-law potential with $\beta<2$,
but does not appear to exhibit tracking behaviour. We have fixed
$\mu=1$ and checked that this does not result in a loss of
generality. For $T_{\rm i}\gtrsim 2$ the solution is very close to de
Sitter, $w_0\approx -1$ and $0<w'_0<10^{-3}$, essentially amounting to
a creeping solution ($T_0\approx T_{\rm i}$) sitting on the asymptotic
flat part of the potential.

For smaller $T_{\rm i}\lesssim 1$ the cosmological values today depend
on the initial conditions of the model. For $T_{\rm i}=1$ we find
$w_0\approx -0.95$, $w'_0\approx 0.04$ and for $T_{\rm i}=0.5$,
$w_0\approx -0.77$, $w'_0\approx -0.05$. The solution exits the
$2\sigma$ contour at about $T_{\rm i}=0.4$ ($w_0\approx -0.70$,
$w'_0\approx -0.14$). One might argue that the initial condition
$T_{\rm i}'=0$ is inappropriate for this model, since the potential is
very steep at small field values. However, we have verified that the
slow-roll velocity is small enough to be negligible anyway at least
for $T_{\rm i}\geq 0.2$ (which is beyond the observable region).
There is no evidence of tracker-type behaviour for $T_{\rm
i}\geq 0.2$.

This model does not need fine-tuning of the initial conditions, as the
field can lie anywhere on the flat part of the potential. However one
has to fix the value of $U_c$ by hand in order to reproduce the
observed dark energy density, which of course does not resolve the
cosmological constant problem.

As in the inverse-power case, we can compare the numerical output of
$w_0$ with the asymptotic solution Eq.~\Eq{-1w}, where
$\phi_0=1/T_0^{\rm num}$. When $\phi_0\sim 1$ and the approximation
giving Eq.~\Eq{-1w} breaks down, $|w_0^{\rm num}-w_0^{\rm
theor}|/|w_0^{\rm num}|\sim 0.1$, while at large $T_0$ one has
$|w_0^{\rm num}-w_0^{\rm theor}|/|w_0^{\rm num}|<10^{-3}$ or
better. We note that Eq.~\Eq{-1wpr} is negative definite while
${w_0^{\rm num}}'>0$, and there is disagreement between the two.


\subsection{$U=U_c\exp(\mu^2T^2)$}\label{exp2}

This model also has a non-zero vacuum energy $U_c$, this time at
$T=0$.  Fixing $\mu=1$, for $T_{\rm i}<0.1$ one has a de Sitter
behaviour ($w_0\approx -1$, $0<w'_0<10^{-3}$), while for increasing
$T_{\rm i}$ the barotropic index goes away from $-1$ (for instance,
$T_{\rm i}=1$ gives $w_0\approx -0.81$, $w_0'\approx 0.13$).

Checking the numerical output of $w_0$ with Eq.~\Eq{2w}, where
$\phi_0=(T_0^{{\rm num}})^2\ll 1$, one can see that $|w_0^{\rm
num}-w_0^{\rm theor}|/|w_0^{\rm num}|<10^{-4}$. Again, Eq.~\Eq{2wpr}
is negative definite while ${w_0^{\rm num}}'>0$.

Note that lower values of $\mu$ lead even more closely to a
cosmological constant behaviour, well inside the $1\sigma$ bound. This
may be relevant when trying to construct a model which fits also for
inflation, as one needs $\mu^2\sim 10^{-8}$ to get the correct level
of anisotropies \cite{GST}.


\subsection{$U=U_c\exp(-\mu T)$ and $U=U_c/\cosh(\mu T)$}

\begin{figure}[t]
\includegraphics[width=8cm]{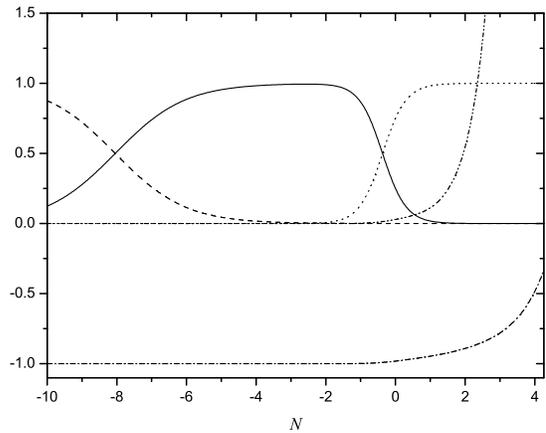}
\caption{\label{fig4} Evolution as a function of $N$ of the model
$U=U_c/\cosh(T)$, in this case extending beyond the present. Solid
line: $\varrho_{\rm m}/x$; dashed line: $\varrho_{\rm r}/x$; dotted
line: $\varrho_T/x$. The dot-dashed and dot-dot-dashed lines are $w$
and $w'$, respectively. In this example, the initial condition is
$T_{\rm i}=0.5$, for which $w_0\approx-0.98$ and $w_0'\approx 0.03$.}
\end{figure}

When considering the pure exponential potential $U=U_c\exp(-\mu T)$,
choosing a different starting value of $T$ has no impact on the
evolution because a rescaling $T\to T+$ const.~simply renormalizes
$U_c$, which the program then adjusts to give the same present status
(corresponding to $w_0\approx -0.93$, $w_0'\approx 0.10$). Therefore
we consider directly the hyperbolic cosine potential of
Ref.~\cite{LLM}, for which this degeneracy is removed. For $0\lesssim
T_{\rm i}\lesssim 0.3$, the solution has $w_0\lesssim -0.99$ and
$0<w_0'\lesssim 10^{-2}$, while for larger values of $T_{\rm i}$ the
present-day barotropic index becomes larger. We note that the
accelerating phase of these solutions is only a transitory epoch
before reaching the dust attractor in the future, as shown in
Fig.~\ref{fig4} where the equations have been integrated up to
positive values of $N$.

Since one must tune, although not too severely, the initial condition
(or, equivalently, $U_c$) so as to get viable acceleration today, this
and other models leading to a dust regime are not very
predictive. They are however capable of explaining the observed dark
energy properties.


\subsection{$U=U_c\exp(-\mu^2T^2)$}

This model has a dust attractor whose qualitative features match the
previous case. It is not difficult to find suitable initial conditions
mimicking a cosmological constant today.


\section{Discussion}

None of the models we have discussed are very satisfactory. Those
which carry reasonable theoretical motivation all end up with a high
degree of fine-tuning. Either the potential normalization $U_c$ has to
be set to match the observed dark energy density, or the initial
conditions tuned to the creeping regime, meaning that the present
density was already set in place during the early Universe. In either
case, this tuning amounts to a restatement of the cosmological
constant problem, rather than a resolution. The models which avoid
fine-tuning of initial conditions (while still subject to tuning of
the normalization), such as the inverse power-law, are constrained by
observations into parameter regimes with no theoretical motivation.

The string tachyon with DBI action seems only weakly competitive as a
dark energy candidate, and another formulation of the effective theory
might have more successful applications. The tachyonic effective
action as the lowest-order level truncation of cubic string field
theory has been studied only very recently and its cosmological impact
has yet to be fully assessed \cite{are04,AJ,cut,AK}. An alternative
would be to abandon the pure four-dimensional picture (ideally
corresponding to a low-energy single or coincident brane configuration
in a higher-dimensional spacetime) and consider more general
brane--anti-brane setups where open string modes naturally
live. However, any modification to Einstein gravity would be severely
constrained after nucleosynthesis.

Finally, it was shown that the addition of electric and magnetic fields in the non-BPS brane world volume can slow down the evolution of the inflationary tachyon and relax the fine tuning of the parameters \cite{CQS,cre05}. It might be worth checking whether the insertion of non-trivial fluxes plays some role in the late universe.

Still, the era of high-precision cosmology opened up by microwave
background and large-scale structure observations is allowing us to
constrain inflationary and dark energy models in a more and more
stringent way, selecting some of them from a plethora of
possibilities. Here we have given an example of the first stage of
such a procedure. The second one will be to develop and refine those
models that seem particularly promising, by embedding them in a
comprehensive and consistent picture of the cosmological history and
its particle theory content.


\begin{acknowledgments}

G. C.\ was supported by Marie Curie Intra-European Fellowship No.
MEIF-CT-2006-024523 and partly by PPARC (UK), and A. R. L.\ by PPARC
(UK). We thank A. De Felice, M. Fairbairn, S. A Kim, E. V. Linder, G. Tasinato, Y. Wang, and especially Pia
Mukherjee for useful discussions and comments.

\end{acknowledgments}


\appendix


\section{Asymptotic solutions}

In this Appendix we derive some analytical asymptotic solutions
describing the evolution once the tachyon has become completely
dominant.

\subsection{Asymptotic solutions for power-law potentials}

Inverse power-law potentials, introduced in the context of dark energy
in Ref.~\cite{RP}, have no general support from string
theory. However, they can be viewed as large-field approximations
($|T|\to+\infty$) of the exact solution \cite{fei02}
\ba
a &=& \exp [p(t/t_0)^n-p]\,,\\
U_n&=&U_c\sqrt{1+T^{2n/(n-2)}}\,T^{-4(n-1)/(n-2)}\label{pot1}\,,
\ea
where we have normalized $a$ so that $a_0=1$. We have
\ba
U_n  &\sim& T^{-\beta}=T^{-4(n-1)/(n-2)},\label{pot2}\\
n    &=&    \frac{2(2-\beta)}{4-\beta}\,,\label{defnb}
\ea
for $0<n<2$, or
\ba
U_n  &\sim& T^{-\gamma}=T^{-(3n-4)/(n-2)},\label{pot3}\\
n    &=&    \frac{2(2-\gamma)}{3-\gamma}\,.
\ea
for either $n<0$ or $n>2$. The solutions with the potential
Eq.~\Eq{pot2} are real, nontrivial and expanding if, and 
only if,
\be
N_*<0\quad{\rm and}\quad 0<\beta<2\quad(0<n<1)\,,\label{att}
\ee
or
\be
N_*>0 \quad{\rm and}\quad \beta<0\quad(1<n<2)\,,
\ee
where $N_*\equiv -p$. In the first case, the big bang event is at
$N_*$ and $N_*<N<+\infty$; in the second case, $-\infty<N<N_*$. The
solutions with the potential Eq.~\Eq{pot3} are well behaved when
$N_*>0$ and either $2<\gamma<3$ ($n<0$) or $\gamma>3$ ($n>2$).

The case given by Eq.~\Eq{att} is the most interesting since it
corresponds to the tracking regime. Using the definition
Eq.~\Eq{defnb} and neglecting matter and radiation contributions, one
has the exact solution
\ba
x  &=& \left(\frac{N_*}{N_*-N}\right)^{\beta/(2-\beta)},\label{ee1}\\
w  &=& -\frac{x'}{3x}-1 = (w_0+1)\frac{N_*}{N_*-N}-1\,,\\
w' &=& -\frac{3(2-\beta)}\beta(w+1)^2\,,\label{wpr}\\
T'^2 &=& \frac{w+1}x\nonumber\\
      &=&
(w_0+1)\left(\frac{N_*}{N_*-N}\right)^{2(1-\beta)/(2-\beta)},\\ 
T &=& 
\frac{\beta}{3\sqrt{w_0+1}}
\left(\frac{N_*-N}{N_*}\right)^{1/(2-\beta)}+C\,,\label{ee5}  
\ea
where we have assumed $T'>0$ ($T>0$), $C$ is an integration constant,
and
\be\label{w0}
w_0+1 = -\frac{\beta}{3(2-\beta)}\frac1{N_*}.
\ee
One can find the value of $p$ for the attractor from today's value
of the barotropic index. Inverting Eq.~\Eq{w0},
\be\label{p}
p=\frac{\beta}{3(2-\beta)(w_0+1)}.
\ee
Also,
\ba
T_0 &=& \frac{\beta}{3\sqrt{w_0+1}}+C,\label{to}\\
T_0'  &=& \sqrt{w_0+1}\,,\label{top}\\
w'_0 &=& -\frac{3(2-\beta)}\beta(w_0+1)^2.\label{wath}
\ea
The classical stability of these solutions was studied in
Refs.~\cite{AF,CGST}. At late times (i.e., large $T$ and $N$), one has
a de Sitter regime for $0<\beta<2$, $\epsilon\to 0$, or an
asymptotically dust solution ($\gamma>2$). These solutions are
actually attractors. As regards the latter, we note that
\be
\dot T^2=(w_0+1)\left(\frac{N_*}{N_*-N}\right)\to 1,
\ee
where we have taken the Carroll limit \cite{gib03} (the DBI action
then becomes singular) corresponding to tachyon condensation into
dust. This expression prescribes a regularization for all the above
formul\ae, and suggests the following possibility, considered also in
Ref.~\cite{CGST}. The only way to balance the increasingly small
denominator is to impose that $w_0+1\approx 0$; this might mean that
if $\gamma>2$, the Universe tends to become dust dominated but only
after passing through an accelerating phase. Since the origin of time
is arbitrary, such a phase is not positioned unequivocally and it will
be determined also by the normalization constant $U_c$ of the
potential.


\subsection{Asymptotic solutions for exponential potentials}

In general, a solution for the Friedmann equation in terms of $T$ can
be found by noting that, when $w\approx{\rm const}$, the $T$
dependence of the Hubble parameter must be the same as of the tachyon
potential, the square root in the denominator of Eq.~\Eq{fr} (with
$\varrho=0$) being dimensionless: $x(T)\propto U(T)$. All the
exponential potentials we consider can be suitably parametrized so
that
\be
x= A \exp[(\mu T)^\alpha]\,,
\ee
where $A=\exp[-(\mu T_0)^\alpha]$ and we have neglected all
matter/radiation contributions. Differentiating this equation with
respect to $N$ and using the continuity equation ($w'=0$), one has
\be
3xT'=-\alpha\mu^\alpha T^{\alpha-1},
\ee
which can be integrated from $N$ to today:
\be
N=-\frac{3A}{\alpha\mu^\alpha}\int_{T_0}^T dT \, T^{1-\alpha} \exp[(\mu
T)^\alpha]\,.
\ee
Defining the variable $\phi\equiv (\mu T)^\alpha$, the integral
becomes 
\ba
N &=& -\frac{3A}{(\alpha\mu)^2}\int_{\phi_0}^\phi
d\phi\,\phi^{2(1-\alpha)/\alpha} e^\phi\\
   &=&
-\frac{3A(-1)^{2/\alpha}}{(\alpha\mu)^2}\left[
\Gamma\left(\frac2\alpha-1,-\phi\right)-
\Gamma\left(\frac2\alpha-1,-\phi_0\right)\right]\nonumber\\
   &\equiv& F(\alpha,\phi)\,,
\ea
where $\Gamma$ is the incomplete gamma function. The cases of interest
are
\ba
F(1,\phi)  &=& -\frac{3}{\mu^2}(e^{\phi-\phi_0}-1)\,,\\
F(-1,\phi) &=&
\frac{3e^{-\phi_0}}{\mu^2}[\Gamma(-3,-\phi_0)-
\Gamma(-3,-\phi)]\,,\nonumber\\\label{g-1}\\
F(2,\phi)  &=&
\frac{3e^{-\phi_0}}{(2\mu)^2}[\Gamma(0,-\phi)-
\Gamma(0,-\phi_0)].\label{g2}
\ea
The first equation implies that
\be
T(N)=T_0+\frac1\mu \ln\left(1-\frac{\mu^2}3N\right)\,.
\ee
In order to find $T$ as a function of $N$ in the other cases,
Eqs.~\Eq{g-1} and \Eq{g2} can be expanded around large or small
$\phi$: for $\alpha=-1$,
\ba
\phi\ll 1:\qquad F(-1,\phi) &\sim&
\frac{e^{-\phi_0}}{\mu^2}
\left(\frac1{\phi^3}-\frac{1}{\phi_0^3}\right)\,,\\    
\phi\gg 1:\qquad F(-1,\phi) &\sim&
\frac{3e^{-\phi_0}}{\mu^2}\left(
\frac{e^{\phi_0}}{\phi_0^4}-\frac{e^\phi}{\phi^4}\right),\nonumber\\
\ea
while when $\alpha=2$,
\ba
\phi\ll 1:\quad F(2,\phi) &\sim&
\frac{3e^{-\phi_0}}{(2\mu)^2}\ln\left(\frac{\phi_0}{\phi}\right)\,,\\
\phi\gg 1:\quad F(2,\phi) &\sim&
\frac{3e^{-\phi_0}}{(2\mu)^2}\left(\frac{e^{\phi_0}}{\phi_0}-
\frac{e^\phi}{\phi}\right).
\ea
One can then numerically invert the above transcendental functions to
get $\phi(N)$ and
\ba
w  &=& -1-\phi'(N)/3\,,\\
w' &=& -\phi''(N)/3\,.
\ea
When $\phi\ll 1$ (which is the regime typical of the dS-attractor 
solutions), one can show that
\ba
w_0  &=& \mu^2e^{\phi_0}
\left(\frac{\phi_0^2}3\right)^2-1\,,\label{-1w}\\ 
w_0' &=& -\frac{4}{3^3}\,\mu^4e^{2\phi_0}\phi_0^7\,,\label{-1wpr}
\ea
for $\alpha=-1$, while
\ba
w_0  &=& \left(\frac{2\mu}{3}\right)^2e^{\phi_0}\phi_0-1\,,\label{2w}\\
w_0' &=& -\frac{(2\mu)^4}{3^3}\,e^{2\phi_0}\phi_0\,,\label{2wpr}
\ea
for $\alpha=2$. Note that $w'$ is always very small in these models.


\subsection{Comparison with numerical solutions}

Having built the asymptotic analytic solutions, one would like to
check whether the numerical behaviour really approaches such solutions
at late times. In particular, we want to compare the numerical points
in the $w_0$--$w_0'$ plane in parameter space, found via
\ba
w_0 &=&(T'_0)^2-1\,,\\
w'_0 &=& 2w_0T'_0\left[3T'_0+\frac{(U_{,T})_0}{U_0}\right]\label{wpr0}\,,
\ea
with the corresponding semi-analytic expressions given by
Eqs.~\Eq{to}--\Eq{wath} for inverse power-law potentials and by
Eqs.~\Eq{-1w}--\Eq{2wpr} for exponential potentials.  The matter
content today still amounts to $25\%$ of the total energy density and
contributes to the cosmological evolution, so we expect a deviation of
the numerical results relative to the asymptotic, pure tachyonic
solution.

There is another source of discrepancy one should take into account,
namely the approximations implicit in the solutions presented in this
Appendix. In the inverse power-law example, while
Eqs.~\Eq{ee1}--\Eq{ee5} are valid for a polynomial potential as in
Eq.~\Eq{pot1}, the numerical model is actually Eq.~\Eq{pot2}, and in
general the two will give a different running of the barotropic index
$w'$, Eq.~\Eq{wpr0}. In Sec.~\ref{power} we find that ${w'_0}^{\rm
num}$ and ${w'_0}^{\rm theor}$ do disagree, but still there is
remarkable agreement between $w_0^{\rm num}$ and $w_0^{\rm theor}$
[the latter comparison is in fact done between $T_0^{\rm num}$ and
$T_0^{\rm theor}(w_0^{\rm num})$].

In the case of exponential potentials, the asymptotic models have
$w\approx{ \rm const.}$ by construction [$x(N)\propto U(N)$] and the
equations of motion are much simplified. Again, $w_0^{\rm num}\approx
w_0^{\rm theor}$ to good approximation (see Secs.~\ref{exp-1} and
\ref{exp2}), while a comparison of the running would require a more
refined treatment.


\end{document}